# *Evolution of the Stress and Strain field in the Tyra field during the Post-Chalk Deposition and Seismic Inversion of fault zone using Informed-Proposal Monte Carlo*

Sarouyeh Khoshkholgh, Ivanka Orozova-Bekkevold and Klaus Mosegaard
Niels Bohr Institute
University of Copenhagen

## Table of Contents

## Abstract

When hydrocarbon reservoirs are used as a CO2 storage facility, an accurate uncertainty analysis and risk assessment is essential. An integration of information from geological knowledge, geological modelling, well log data, and geophysical data provides the basis for this analysis. Modelling the time development of stress/strain changes in the overburden provides prior knowledge about fault and fracture probability in the reservoir, which in turn is used in seismic inversion to constrain models of faulting and fracturing. One main problem in solving large scale seismic inverse problems is high computational cost and inefficiency. We use a newly introduced methodology - Informed-proposal Monte Carlo (IPMC) - to deal with this problem, and to carry out a conceptual study based on real data from the Danish North Sea. The result outlines a methodology for evaluating the risk of having sub-seismic faulting in the overburden that potentially compromises the CO2 storage of the reservoir.

## Introduction

Sequestration of $CO_2$ in former oil and gas reservoirs can contribute to amelioration of the global increase in $CO_2$ emissions, and it is already in use in a limited number of sites worldwide (Bachu, 2008; Michael et al., 2010; Ringrose et al., 2017). Injection of $CO_2$ for enhanced oil recovery has already been utilised by the oil industry for decades, particularly in onshore North America (Gozalpour et al., 2005). $CO_2$ is currently stored offshore Norway, in Sleipner and Snøhvit, with ~ $1.5 \cdot 10^6$ tonnes annually (Eiken et al., 2011). Pilot-projects have been carried out in Germany (Kempka & Kühn, 2013; Bergmann et al., 2016), Spain (Vilamajó et al., 2013; Ogaya et al., 2013) and Texas (Daley et al., 2008; Doughty et al., 2008), and this has greatly increased our understanding of $CO_2$ migration, monitoring and injection strategies in geological reservoirs.

Injection of $CO_2$, or any other fluid, into subsurface reservoirs might increase the risk of caprock failure and migration of fluid along faults, fracture corridors, and other pre-existing weak zones (Ogata et al., 2014a). Understanding the detection thresholds of such structures calls for careful mechanical modelling of the reservoir stress- and strain field, careful inversion of available seismic data, combined with geological knowledge from well data and outcrop data. In this way it may be possible to quantify the probability of significant $CO_2$ migration through the caprock, laying the ground for a meaningful risk evaluation.

During the last couple of decades considerable progress has been made in geophysical and geostatistical data analysis methods to correctly estimate model uncertainties and thereby to evaluate fault detection thresholds (zee, e.g., Zunino et al., 2015). The goal of this pilot project is to propose a way of exploiting these methods for risk assessment in connection with $CO_2$ storage. Conceptual models are developed to model the time evolution of the subsurface, giving information about current and future stress fields resulting from geological processes, such as sediment deposition for example. This analysis provides the prior information to a subsequent probabilistic inversion of seismic data. Monte Carlo methods are used to simulate the noise in the data, and the noise is back-propagated through the geophysical (e.g., seismic) equations into the geophysical model, generating a model variability, and reflecting the uncertainty of the reservoir structure. Combining this approach with prior information about the mechanical properties of the reservoir, we evaluate the probability of fault migration scenarios. In this pilot project, we carry out a concrete, highly simplified numerical study of the sub-problem of estimating the density of sub-seismic faults in the overburden of an existing North Sea hydrocarbon reservoir, and established a simple probability model for releases through existing faults. The study is a starting point for developing a full-scale risk analysis system based on the principles outlined above.

## Well data

Since the overburden provides the reservoir seal and hosts significant part of the infrastructure, it is important to have sufficient data of good quality in order to analyse its integrity and/or strength. One very important parameter is the fracture pressure of the seal – if the pressures at the top of the reservoir exceeds the fracture pressure of the seal a breach will occur and the reservoir fluid ($CO_2$ or hydrocarbons) will escape.

Data (petrophysical logs and well reports) for the following wells: Fasan-1, Deep Adda, South East Adda and E1-X (Figure 1), were provided by DHRTC (Danish Hydrocarbon Research and Technology Centre). The wells are located in the Tyra field, in the Danish sector of the Central Graben, approximately 200 km west from the city of Esbjerg.

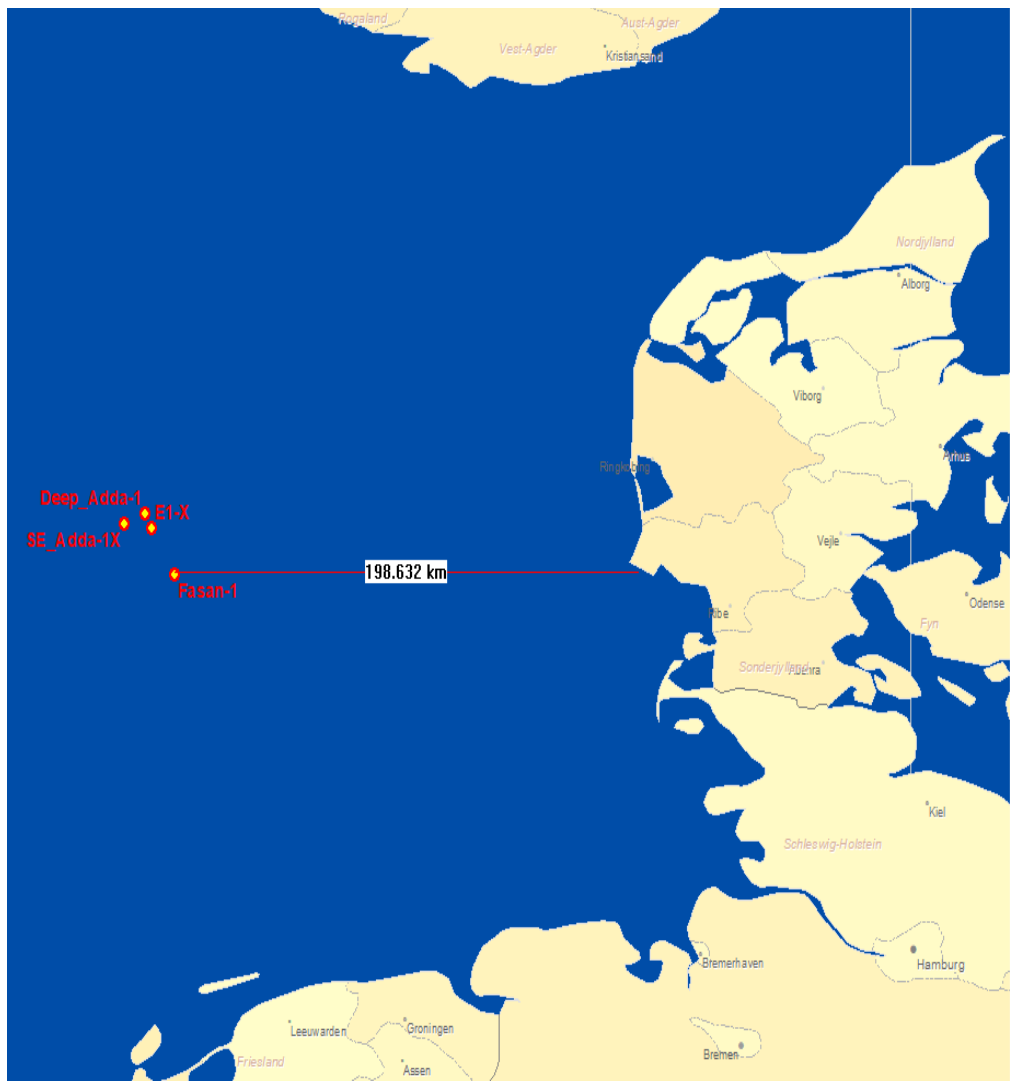

**Figure 1.** Location of the wells used in the analysis.

Figure 2 and 3 show the lithology columns and, respectively, the depth coverage of the sonic and density logs acquired in the four above mentioned wells. The figures illustrate very well the challenges related to the availability of log data in the overburden. In all four wells, no sonic (or density) data were acquired in the shallow section from sea bed to approximately 500 m depth. In the well E-1X, both sonic and density logs were acquired only in the reservoir (chalk) section, while in South East Adda, density data are available only below the Chalk group, in the deepest section of the well. This probably is due to the fact that the purpose of the South East Adda well was to investigate the hydrocarbon potential of the Cromer Knoll Group (Lower Cretaceous).

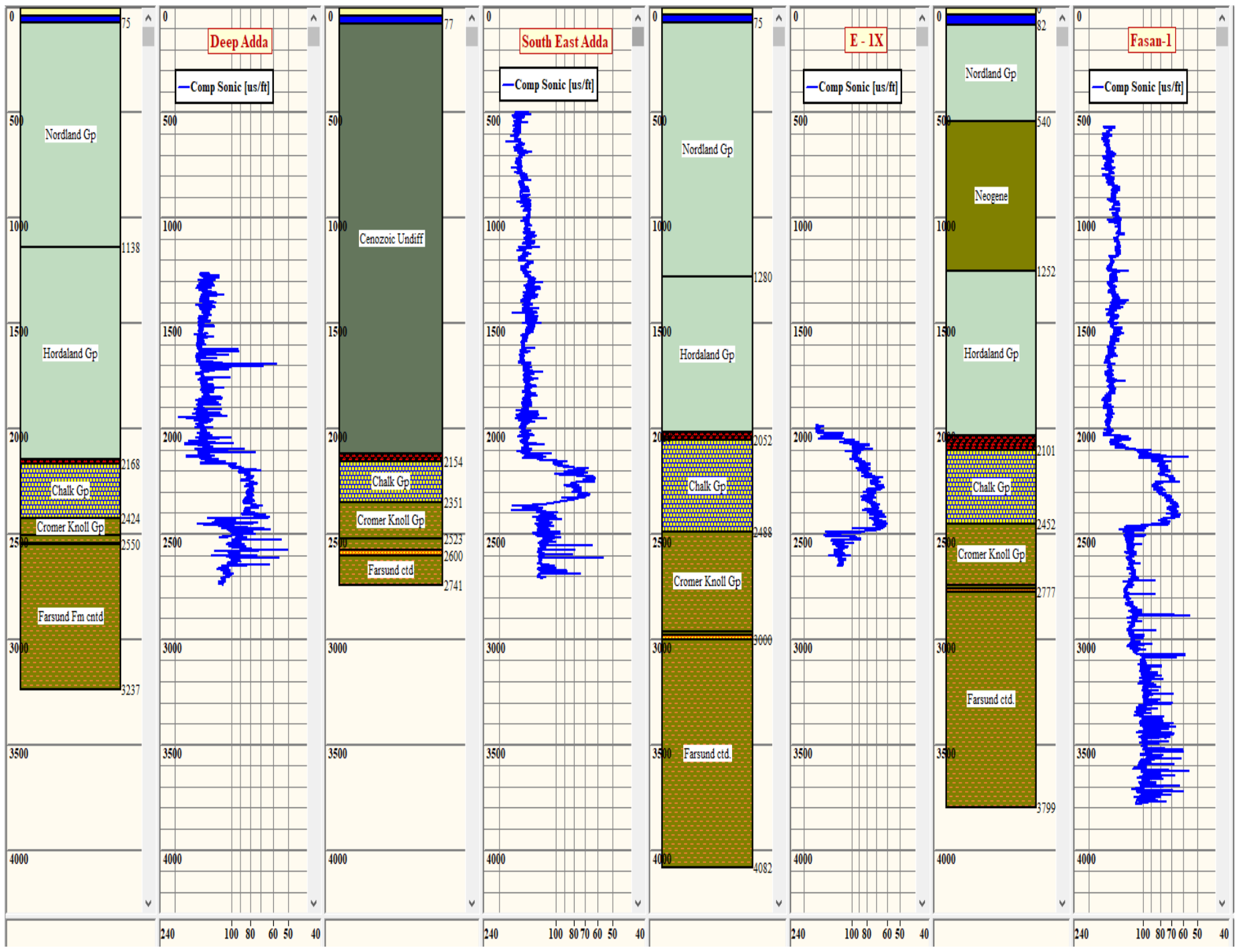

**Figure 2.** Overview of the sonic compressional logs (us/ft), acquired in the four wells. From left to right: Deep Adda, South East Adda, E-1X and Fasan-1. The depth reference is Rotary Table (RT, i.e. the rig floor).

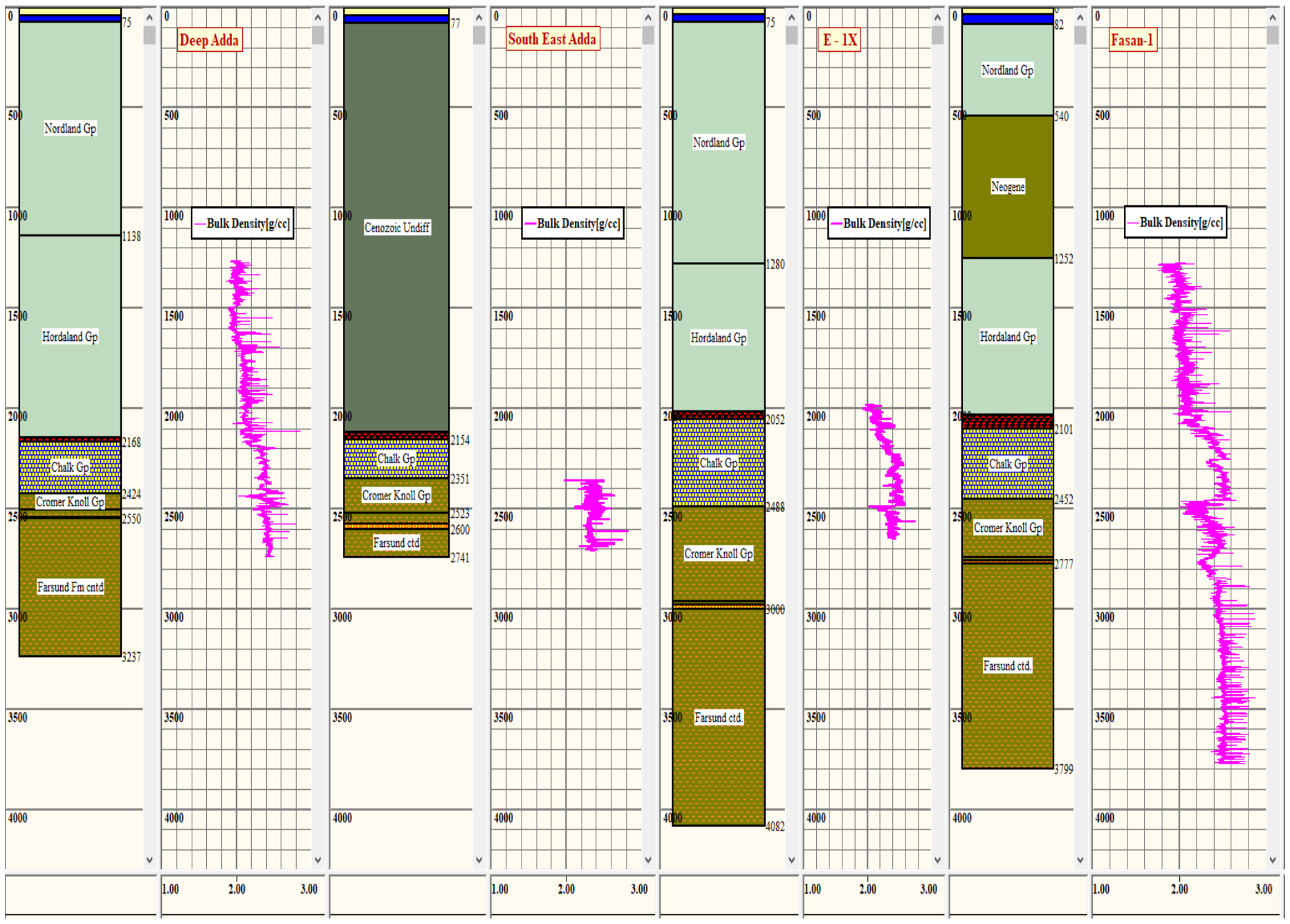

**Figure 3.** Overview of the density logs (g/cc), acquired in the four wells. From left to right: Deep Adda, South East Adda, E1X and Fasan-1. The depth reference is Rotary Table (RT, i.e. the rig floor).

## Modelling of the the Post-Chalk overburden in the Tyra Field

The hydrocarbon field Tyra is situated in the Danish sector of the Central Graben in the North Sea. The hydrocarbon accumulation in this field is mainly concentrated in the Chalk Formations Ekofisk (Danian), Tor and Hod (both Upper Cretaceous).

The present-day stress-strain state of the subsurface is the result of complex geological and planetary processes, taking place over millions of years, such as plate tectonic movements, sedimentary deposition, climate changes, erosion, uplifting etc.

The most dominating geological process in the Post-Chalk period, i.e. the last 61 Million Years (MY) from Mid Paleocene (Selandian) to Present, was the filling in of the North Sea Basin with sediments originating from Fennoscandia and Scotland (Konradi, 2005, Schiøler et al., 2007, Gibbard and Lewin, 2016). The most intensive sedimentary flux into the North Sea occurred during the Late Pliocene-Pleistocene (Gibbard and Lewin, 2016 ). These sediments, deposited in offshore environment (Schiøler et al., 2007), form the present day overburden above the chalk reservoirs in the North Sea.

Several phases of inversion and compression in the North Sea during the Paleocene (Nielsen *et al.*, 2005; Japsen et al., 2014)  and in the Early Miocene (Rasmussen, 2004) have been suggested. However, it is believed that the magnitude of these events, especially the Miocene one, did not affect significantly the depositional process in the Central Graben.

Thus, the present-day stress-strain state of the Post-Chalk overburden in the North Sea is mainly the result of the sedimentary deposition. In addition, the weight of the thick overburden affects also the stress-strain field of the underlying strata.

As mentioned above, the overburden provides the reservoir seal, and its integrity is a crucial parameter to consider when evaluating the risk of leakage of reservoir fluid ($CO_2$ or hydrocarbons).

Reservoir depletion, caused by hydrocarbon production, leads to stress and strain changes both in the reservoir and the overburden as well, among the examples are the subsidence experienced in the Tyra (Plischke, 1994; Schutjens et al., 2019) and the Ekofisk (Sulak and Danielsen, 1989) fields.

In the Tyra field, the Post-Chalk overburden is approximately 2000 m thick and is represented by three main groups: the thick Nordland and Hordaland Groups, composed mainly of smectite-rich shales (Nielsen and Rasmussen, 2015), and the much tinner Rogaland Group (Figure 2 and 3).

The well reports of the four wells (Jensen, 2004; Rong et al., 1985; Kleist et al., 1977), used in this study, hold additional useful information about the lithological composition of the overburden. The upper (most shallow 0-500 m depth) part of the Nordland Group consist of predominantly Quaternary sand sand/clay mixtures, while the lower part consist predominantly of claystone with occasional thin limestone layers (Jensen, 2004). The Hordaland Group consists predominantly of clay-rich (shale) formations, in some interval interbedded with thin limestone layers (Jensen, 2004). The Rogaland Group is situated at the top of the Chalk Group and thus represents the seal for the uppermost Chalk reservoirs (Danian age, Lower Paleocene). The thickness of Rogaland varies across the field, as illustrated by the well lithology columns in Figure 2 and 3. The upper part of the Rogaland Group (the Balder formation) is characterized by tuffaceous claystone, while the lower part (Sele, Lista, Vaale) is predominantly claystone with stringers of marl (Jensen, 2004).

The Rogaland Group, which is overlaying the Danian Chalk sequences, spans the period from Selandian (Mid Paleocene) to Ypresian (Early Eocene), approximately from 61 to 53 million years ago (Ma); the Hordaland Group spans the period from Lutetian (Mid Eocene) to Tortonian (Mid/Late Miocene), from 53 to around 7 Ma; and the Nordland Group spans the period from Messinian (Late Miocene) to Present Day, roughly the last 7 million years. (Lithostratigraphic Chart of the Central North Sea, 2014, https://www.npd.no).

## Numerical representation of the Post-Chalk Deposition in the Tyra Field

A 2D subsurface geometry in the Tyra Field was derived from seismic interpretation of the main reflective horizons down to (approximately) 2500 m depth below sea bed.

These main reflective horizons and the corresponding stratigraphic units are shown in Figure 4. The bottom of the domain was set to 2500 m below sea. The lateral (horizontal) extension is approximately 2000 m. It was assumed that the horizons correspond roughly to the stratigraphic groups/formations and the geological age given in the Table 1. The overburden section is composed by the strata above the Chalk Group (given in blue on Fig.4, bottom right).The steep sections of the horizons representing the tops of the Jurassic and the Lower Cretaceous sections probably indicate the presence of faults.

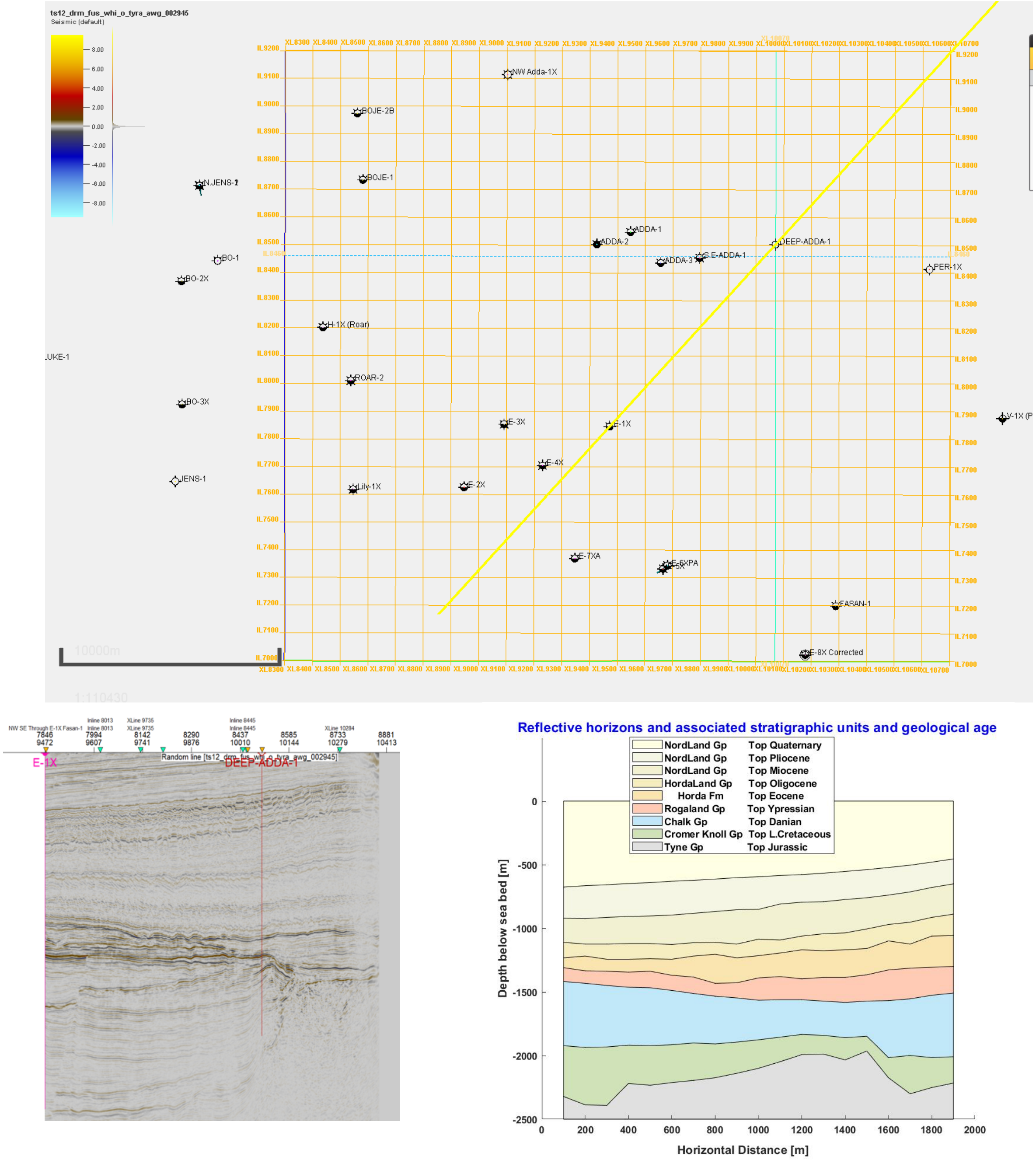

**Figure 4.** Sketch of the location of the considered seismic profile, the seismic data, and the main reflective horizons, derived from seismic interpretation and the corresponding stratigraphic units and geological age. The steep sections of the horizons representing the tops of the Jurassic and the Lower Cretaceous sections probably indicate the presence of faults.

In the present model, the overburden was deposited upon a pre-existing domain (called also underburden), composed by the strata from Top Jurassic to Top Danian, Fig. 4. Thus the pre-existing domain is composed by the units (Fig. 4) Tyne (grey), Cromer Knoll (green) and Chalk (blue).

The horizons from Top Ypressian to Top Quaternary (Fig.4, Table 1) , were used to define the depositional stages in the time evolution of the Post-Chalk period.

| Horizons, from Fig. 4 | Stratigraphic Group / Formation | Numerical Age (Ma) |
|---|---|---|
| Sea Bed / Top Quaternary | Top Nordlang Gp | 0.00 |
| Top Pliocene | Mid Nordland Gp | 2.60 |
| Top Miocene | Lower Nordland Gp | 5.30 |
| Top Oligocene | Upper Hordaland Gp /Mid Lark Fm | 23.00 |
| Top Eocene | Lower Hordaland Gp/ Top Horda Fm | 33.90 |
| Top Ypresian (Mid Eocene) | Top Rogaland Gp | 47.80 |
| Top Danian (Early Eocene) | Top Chalk / Top Ekofisk Fm | 61.60 |
| Top Upper Cretaceous | Top Shetland Gp / Top Tor Fm | 66.00 |
| Top Lower Cretaceous | Top Cromer Knoll Gp | 100.0 |
| Top Jurassic | Top Tyne Gp | 145.0 |

**Table 1.** Tentative assignment of the reflective horizons to stratigraphic groups an geological age.

The post-Chalk deposition was modelled using the approach, recently presented by Orozova-Bekkevold et al. (2021). Each stage is defined by specific duration in millions of years and a prescribed number of discrete depositional layers, composing the respective section. The duration in time and the number of discrete layers used for each depositional stage are summarized in Table 2.

Since the focus of this study was the Post-Chalk overburden, the geological evolution of the Jurassic and the Cretaceous sections pre-existing base (sedimentary deposition, tectonic events, erosion, uplifting, etc.) was not modelled in details at this stage. In addition we lacked data related to material properties and timing of past tectonic events for these sections.

| Stage | Geological Period | Duration [MY] | Nr of layers |
|---|---|---|---|
| 1: Pre-existing base | Upper Jurassic | 0.1 - Settling under gravity | 1 |
| 2: No detailed modelling | L. Cretaceous | 1.0 - Settling under gravity | 1 |
| 3: No detailed modelling | U. Cretaceous – Danian | 1.0 - Settling under gravity | 1 |
| 4: | Ypresian (L. Eocene) | 8.0 | 4 |
| 5: | Mid to Late Eocene | 19.3 | 5 |
| 6: | Oligocene | 10.7 | 5 |
| 7: | Miocene | 17.7 | 5 |
| 8: | Pliocene | 2.7 | 2 |
| 9: | Quaternary | 2.6 | 2 |
| 10: | Post-depo settling | Projected 0.5 MY after present | Not used |

**Table 2.** Main deposition stages with duration (in Million Years, MY) and present day thickness.

The layers were deposited one-by-one on the top of the underlaying ones until the height of the respective horizon (Fig. 4) was reached. At the end, after the deposition of the last layer, a ‘settling’ period of the duration of 0.5 MY was introduced to simulate conditions of relaxation with no deposition.

The Jurassic and Cretaceous sections were modelled as "deposited" in one chunk and were "allowed" to settle under gravity, before the onset of the Post-Chalk deposition. The material, composing the Upper Jurassic (stage 1 in Tab. 2) is assumed to be consolidated shale, while the Low Cretaceous to Top Danian (stage 2) is assumed to be composed by consolidated sandstones.

The stratum between Top Lower Cretaceous and Top Danian can be seen as a proxy for a reservoir.

It is assumed that the deposited material composing the Post-Chalk overburden is unconsolidated clay and all materials are considered to be

- isotropic and homogeneous;
- fully saturated with water;
- and at any given time the sediments are at their maximum burial depth.

Clay diagenesis and other chemical and temperature effects are not modelled at the current stage.

## Finite elements modelling of the Post-Chalk deposition in the Tyra Field

The evolution in geological time is modelled in the terms of a finite element method, using the software Elfen (ELFEN, Rockfield Software Ltd.). The framework and the theory behind the software are given in details in Crook et al. (2003), Peric and Crook (2004), (Crook et al. (2006a, 2006b), Thornton and Crook (2014).

The non-consolidated clay material composing the overburden is represented as a poro-elasto-plastic material, fully saturated with water. The mechanical field (the solid part) is solved explicitly, while the seepage field is solved by implicit time integration schemes and the two fields are coupled at given time intervals (Thornton and Crook, 2014).

The mechanical properties of the water-saturated medium, were expressed as:

$$div(\boldsymbol{\sigma}')+[(1-\varphi)\rho_s+\varphi \mathrm{P}_f](\mathbf{g}-\boldsymbol{a}_s)=0\,. \qquad (1)$$

The fluid (water only) flow over geological time is represented with a transient equilibrium equation:

$$div\left[\frac{\mathrm{k}(\varphi)}{\mu}\nabla\mathrm{P}_f-\rho_f(\mathbf{g}-\boldsymbol{a}_s)\right]=[\varphi/K_f+(\alpha-\varphi)/K_s]\frac{\partial\mathrm{Pf}}{\partial\mathrm{t}}+\alpha\frac{\partial\varepsilon\mathrm{v}}{\partial\mathrm{t}} \qquad (2)$$

where:

$\boldsymbol{\sigma}' = \boldsymbol{\sigma}-\boldsymbol{\alpha}\mathbf{P}$ is the effective stress; $\rho_s$ and $\rho_f$ are the solid and the fluid density, respectively; **g** is the Earth's gravitational acceleration, $\mathbf{a}_s$ is the acceleration of the solid phase, $\mathbf{P}_f$ is the fluid pressure, $\boldsymbol{\varphi}$ is the porosity, **k(φ)** is the porosity-dependent permeability, $\mathbf{K}_f$ is the fluid bulk modulus, $\mathbf{K}_{fr}$ is the frame bulk modulus, $\boldsymbol{\alpha}$ is the Biot's coefficient and $\boldsymbol{\varepsilon}_v$ is the volumetric strain.

The bulk modulus of the non-consolidated clay is expressed as a function of the mean effective stress P' (Thornton and Crook, 2014):

$$K = Ko + \frac{(1-A)Pco}{\kappa} exp\left[\frac{\varphi_O - \varphi}{\lambda(1-\varphi_O)(1-\varphi)}\right] + \frac{A*\sigma'}{\kappa(1-\varphi)} \qquad (3)$$

where
$\varphi_o$ is the initial porosity, $K_o$ is the initial bulk modulus and $P_{co}$ is the initial pre-consolidation pressure, A is a weighting factor, $\kappa$ and $\lambda$ are material constants.

The vertical stress, $S_V$, resulting by the weight of the overlaying sediments is assumed to be the maximum stress. The horizontal stresses is assumed to be isotropic, and in the absence of tectonic forces, it is derived from the vertical stress:

$$S_h = K_{eff} * S_V \qquad (4)$$

where $K_{eff}$ is the so-called effective stress ratio (Matthews and Kelly, 1967). Finally, the mean effective stress is obtained as

$$\sigma' = \left[\frac{Sv+2*Sh}{3}\right] - \alpha P_p \quad = \sigma - \alpha P_p \qquad (5)$$

Where is $P_p$ the pore pressure and $\alpha$ is the Biot's coefficient. Density and porosity are derived by well data (sonic and density logs), following the procedure presented in Orozova-Bekkevold et al. (2021).

The evolution of these properties in time is driven by the rate of deposition and subsequent burial of fully saturated material, as described in Orozova-Bekkevold et al. (2021), Crook et al. (2003), Peric and Crook (2004), Crook et al. (2006a, 2006b), Thornton and Crook (2014).

***Initial and boundary conditions***: The main force acting upon the domain is the gravity. The gravitational load originates at the top of the sediments and acts downwards. This setup is considered representative for the Cenozoic Period in the North Sea basin, where no major tectonic events (uplift, erosion, collision, subduction etc.) occurred, and thus the maximum stress is caused only by the weight of the deposited material and acts in the vertical direction. Uniaxial compaction (i.e. plain strain conditions) is assumed.

The domain below the overburden (Jurassic to Danian, stage 1-3 in Tab. 2) is not allowed to deform at the bottom, along the top and across the sides.

As mentioned above, the deposited material is modelled as a fully water saturated porous medium. The formation water can flow both vertically and horizontally within the domain, but there is no fluid flow from outside sources. The water does not flow through the bottom and no capillary and temperature effect are taken into account at this stage.

***Meshing:*** The finite element mesh is generated by an advancing front algorithm, adding new elements as the geometry expands, following the deposition of new layers of material (Crook et al. (2003), Peric and Crook (2004), Crook et al. (2006a, 2006b)). The elements are triangular, with initial size of 100m. The size of the elements is rescaled during the simulation, depending on the estimated plastic strain at a given step: plastic strain exceeding 2 results in diminished element size.

At the beginning of the simulation, before start of the deposition, the finite elements mesh consist of around 120 triangular elements, all with size 100. At the end of the simulation, the mesh consists of around 6750 elements with size ranging from 50 to 100. The model geometry and the finite element mesh at the beginning and the end of the deposition is summarized in Figure 5.

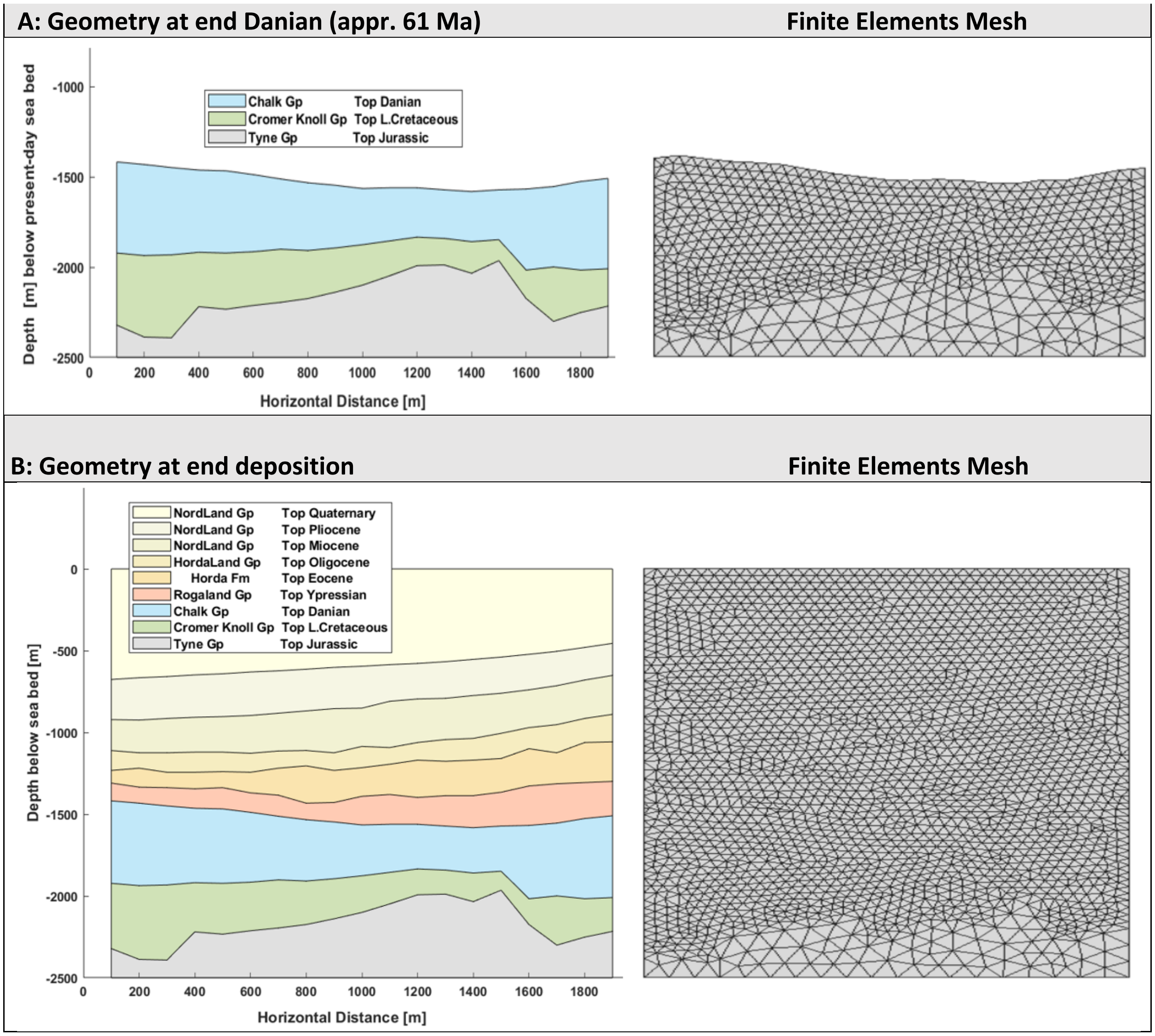

**Figure 5.** Model Geometry and the finite elements mesh in time. The depth is given meters [m] below sea bed at present. From top to bottom: a) Pre-existing domain (underburden) at time appr. 61 Ma (end Danian) before the onset of the overburden deposition; b) The final stage at present time after appr. 61 MY of deposition.

## Results: Evolution of the stress and strain state in the Tyra Field during the Cenozoic Deposition

The estimated changes in time in the subsurface stress-strain fields as result of the overburden deposition in the last 61 MY are reported in this section.

If the shear (differential) stress exceeds the strength of the material, it might result in fracturing of the formation. Figure 6 shows the evolution of the shear stress during the deposition of the overburden. The plot in the centre of the top row (labelled “Top Chalk – 61 Ma”) corresponds to the stage just before the onset of the Cenozoic deposition, approximately 61 Million years Ago (Ma). As the figure shows, the largest shear stress was estimated along the very steep shoulders of the deepest horizon, which corresponds to the boundary between the Upper Jurassic and the Lower Cretaceous. Relatively high stress magnitudes were also estimated in the lowest strata, roughly corresponding to the Chalk reservoirs, while the shear stress estimated in the overburden was low. It is interesting to observe that in the 0.5 MY after the end of deposition, the projection suggests that the subsurface continues to react and a re-distribution of the shear stress might occur, especially in the deepest layers (right bottom corner of the plot).

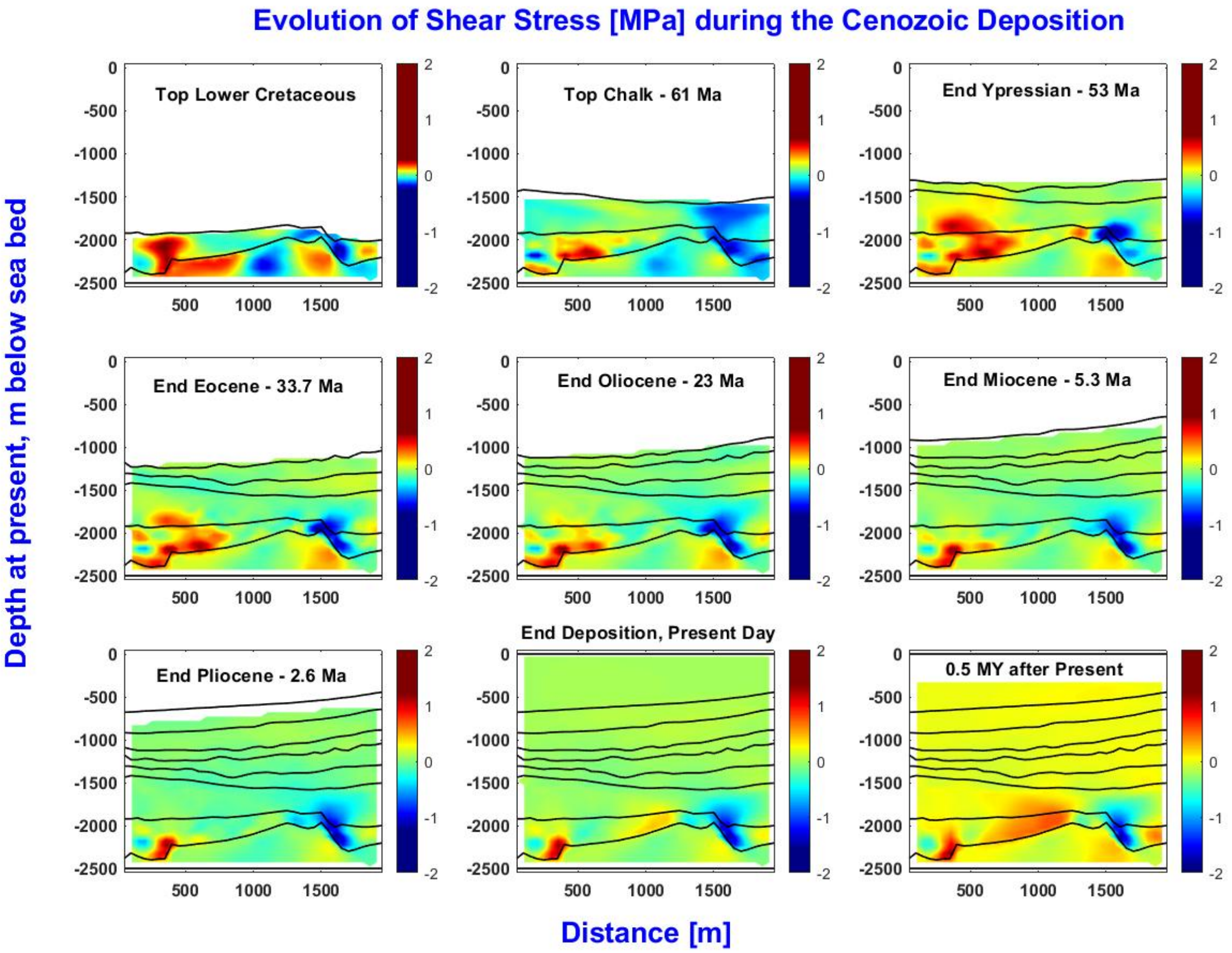

**Figure 6**. Evolution of the Shear Stress [MPa] during the deposition of the overburden. Depth is in meters, below sea bed. The horizons are at their present day-depth. Shear Stress is given in MPa

The zones with large shear stress might be considered being at higher risk of fracturing. The largest shear stress magnitude are estimated at present time where the overburden build-up is completed. If no further deposition is going to occur in the next 0.5 MY, the medium relaxes and the magnitude of the shear stress could decrease.

Figure 7 shows the evolution in the effective strain during the burial process. This could also be interpreted as a kind of deformation accumulated during the Post-Chalk sedimentation. The maximum values were estimated in the overburden, along the Danian-Ypressian boundary, in the strata which forms the top seal of the Upper Chalk reservoirs. The strain continues to increase also in the projected post-deposition period (the right bottom plot in Fig. 7).

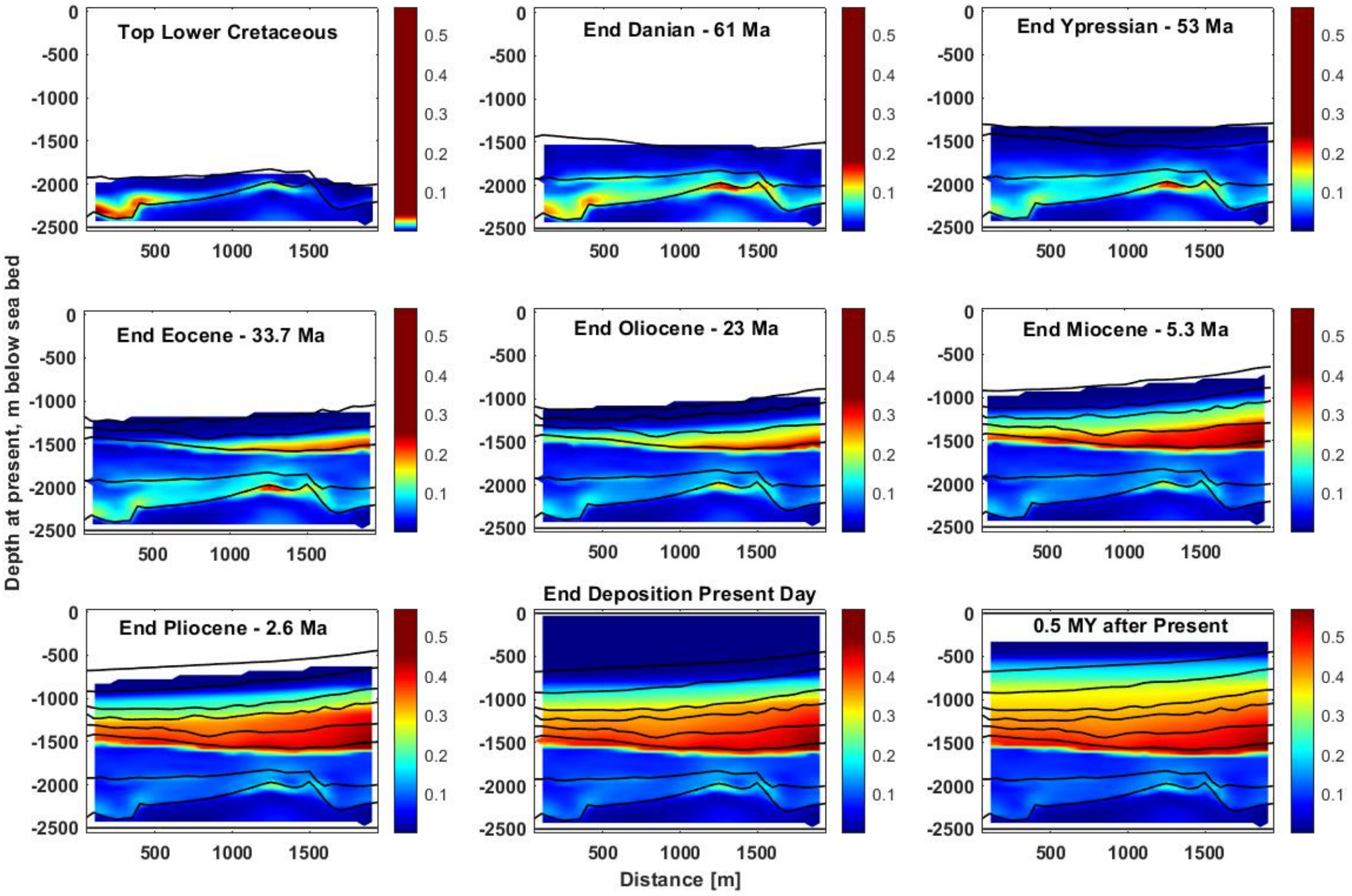

**Figure 7**. Evolution of the effective strain as result of the deposition of the Post-Chalk sediments.

## Seismic Study of the fault Detection Limit

Seismic reflection data is used extensively in the oil and gas sector for imaging and characterizing the subsurface. Reflected energy and arrival time of reflected waves provides information on elastic properties in specified locations and allows us to map geological formations on large and smaller scales, only limited by the resolution of the data.

Seismic resolution implies the capability of distinguishing between different geological micro-structures. Some geological features such as small faults and fractures are below the seismic resolution threshold. Recognition of small-scale structures helps us to study thin layers and minor fractures and faults in reservoirs for future exploration or risk analysis (Ashraf, 2020). When conducting risk assessment in the area, the fault detection threshold in the model, or the smallest fault size that can be detected by seismic data, as well as prior knowledge of the mechanical properties of the rock, such as the stress and strain condition of the overburden, must be taken into account.

Using probabilistic models for evaluating the fault density in a certain area, we can assess the likelihood of CO2 migration in the reservoir. One of the first investigations of possible leakage risks and safety of CO2 underground injection was conducted by Holloway (1997), and probabilistic methods for risk assessment of such problems has become more and more popular during years (Kopp et al., 2010; Smith et al., 2011; Zunino et al., 2015 ).

In this study we use a simplified probabilistic approach to locate areas of increased risk of fracturing in the cap rock. We use a Monte Carlo method to generate small subseismic faults escaping the resolution limit of the data, but at the same time having a fault density proportional to a prior probability density derived from the differential stresses presented in the previous section.

Markov Chain Monte Carlo algorithms are commonly used for solving nonlinear inverse problems and sampling posterior probability density where the aim is integrating independent sources of information such as geological information and prior knowledge, geophysical data and data from nearby well logs. In stochastic approaches such as the Monte Carlo method, the size of the model space and high calculation cost associated with forward modelling could be challenging. Khoshkholgh et al., 2021 proposed a new methodology known as Informed Proposal Monte Carlo (IPMC), for MCMC sampling method that uses external information obtained from simplified physics to form a global proposal distribution that guides the sampling procedure. An example of solving a probabilistic Full-waveform inversion by IPMC for a near surface velocity model in large scale can be found in Khoshkholgh et al., 2021.

In this study we use Informed Proposal Monte Carlo in order to integrate (1) prior knowledge about strain and stress fields and fault and fracture probability models in the reservoir, and (2) information from observed seismic data, to provide uncertainty analysis and to assess the probability of CO2 migration. The aim is to contribute to probabilistic models that could appraise possible CO2 leakage through faults in the overburden.

## Seismic Inversion

In our inversion approach, our aim is to produce subsurface models presenting relevant physical characteristics of the caprock in the chosen area. If $\boldsymbol{m}$ denotes the model parameters (in our case acoustic impedance), $\boldsymbol{d}$ the data and $g$ stands for the forward operator (represented by the wave propagation algorithm), the forward problem can be expressed as:

$$\boldsymbol{d} = g(\boldsymbol{m}).$$

The inverse problem, in its widest sense, can be described as a problem of integrating measurable data information, prior information and theoretical connections between models and data (Tarantola et al. 1982). As mentioned by Tarantola and Valette (1982) the solution of the inverse problem is then defined as a posterior probability density which is obtained as a product of the prior probability density $\rho$, providing the probability of models based on non-seismic information, and the likelihood function $L$, measuring the degree of fit between the observed data, and synthetic data calculated from the elastic reservoir model:

$$\sigma(\boldsymbol{m}) = k\,\rho(\boldsymbol{m})\,L(\boldsymbol{m}).$$

$k$ is the normalization constant and $\sigma(\boldsymbol{m})$ denotes the posterior probability density that is considered as the solution of inverse problem. In our case, $\rho(\boldsymbol{m})$ is derived from a simple hypothesis about proportionality of fracture risk and shear (differential) stress, and $L(\boldsymbol{m})$ is given by

$$L(\mathbf{m}) = \mathrm{C}\exp\left(-\frac{1}{2}(\boldsymbol{d}_{obs} - g(\boldsymbol{m}))^T \boldsymbol{C}_D^{-1}(\boldsymbol{d}_{obs} - g(\boldsymbol{m}))\right)$$

where $\boldsymbol{d}_{obs}$ are the (noisy) observed data, and $\boldsymbol{C}_D$ is the covariance matrix of the noise. In this study we assume that $\boldsymbol{C}_D = \sigma_D^2 \boldsymbol{I}$, where $\sigma_D$ is the standard deviation of the seismic noise.

When the posterior probability density cannot be calculated analytically, Monte Carlo sampling is required (Mosegaard, 2006). The posterior distribution can be described by an ensemble of realizations implemented by Markov-Chain Monte Carlo (MCMC) algorithms, and their variability expresses the uncertainty of subsurface structures (Hastings 1970; Metropolis et al. 1963; Mosegaard and Tarantola 1995; Tarantola, 2005). This requires the ability of calculating likelihood $L$ at each point, as well as an algorithm that can sample the prior $\rho$ . When the prior model is complex, as in this study, sampling through numerical operations is often the only way to incorporate this information (Mosegaard 1998; Zunino et al. 2015). In this study our geological prior information comes from two main sources: (1) the shear stress analysis presented in the previous section, providing a fault risk map, and (2) the simplifying assumption that the reservoir zone consists of a stack of (non-horizontal) homogeneous layers whose properties are calibrated to well information in the area (the wells E-1X and Deep Adda).

There are several studies looking at the properties of fault and fractures zones in relation to hydrocarbon investigation or CO2 storage (Aydin, 2000 ; Rotevatn et al., 2011; Shipton et al., 2004; Dockrill et al., 2010). A fault displacement causes two damage zones on both sides, which could be categorized into three different groups: along fault, around tip and cross fault (Choi et al., 2015). The thickness of the damage zone is influenced by the size of the fault displacement. In our prior model generator, fault models in a specific angle range have been proposed in overburden and the fault density is proportional to the shear stress.

Khoshkholgh et al., 2021 proposed a warping strategy for generating prior models that works by warping and deforming the subsurface image and perturbing the velocity value at each layer. This is done by finding a random window inside the model and allowing deformations such that they are maximum in the center and gradually fade away and become zero at the boundaries. The velocity perturbation within one layer is adjusted by an estimated modelization error (Khoshkholgh et al., 2021). In this study, the same technique is applied in such a way that fault planes are randomly (according to the prior) introduced in overburden during perturbation and the displacement vector directions on either side of the fault line are opposite.

We use the seismic convolutional model as the forward model, which produces the seismic data by convolving the reflectivity series with a wavelet. The reflectivity is computed from the seismic impedance, which is a product of density and velocity. We assume here that the density is roughly proportional to the velocity (Liner, 1999). In order to simulate the residual, horizontal smearing left in the data after (an unavoidable imperfect) migration, we use a 2D wavelet (see Figure 8):

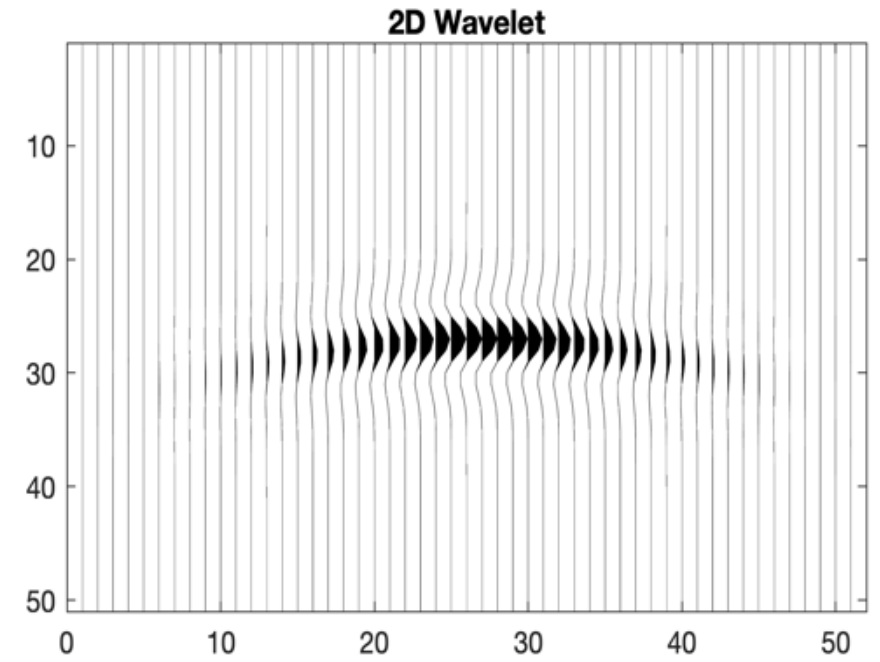

Figure 8. 2D wavelet used in this study

where the variation in time is a zero phase Ricker wavelet. The following is the formula for convolutional modeling of the seimic profile:

$$R(x,t) \ * \ W(x,t) + \ n(x,t) \ = \ D(x,t)$$

Where $R(x,t)$ is the 2D reflectivity, $W(t)$ is the 2D seismic wavelet and $n(x,t)$ is the added noise (Partyka, 1999). For the sake of simplicity, we will assume in this study that the wavelet is constant everywhere in the depth domain, and hence the modeling can be carried out with $t$ replaced by depth $z$.

To carry out the inversion we us a Markov Chain Monte Carlo algorithm which proceeds as follows:

- From current starting model $\boldsymbol{m}$, propose $\boldsymbol{m}'$ using the proposal distribution $q(\boldsymbol{m}' \,|\boldsymbol{m})$
- Accept $\boldsymbol{m}'$ with probability:

$$p_{acc}^{(1)} = min\left(\frac{\rho(\boldsymbol{m}')q(\boldsymbol{m}\,|\boldsymbol{m}')}{\rho(\boldsymbol{m})q(\boldsymbol{m}'\,|\boldsymbol{m})}, 1\right)$$

  where $\rho(\boldsymbol{m})$ is the prior probability density.

- If $\boldsymbol{m}'$ is accepted, perform another test, where it is accepted with probability

$$p_{acc}^{(2)} = min\left(\frac{L(\boldsymbol{m}')}{L(\boldsymbol{m})}, 1\right)$$

  where $L(\boldsymbol{m})$ is the likelihood function.

- If $\boldsymbol{m}'$ is rejected by both of the above tests, repeat $\boldsymbol{m}$.

In order to construct an MCMC algorithm with an informed proposal, we first assume that data uncertainties are very small in comparison to the modelization error (Khoshkholgh et al. 2021). This results in the following formula:

$$\sigma_m(\boldsymbol{m}) \propto \sigma\,(\boldsymbol{d_{obs}}, \boldsymbol{m}) \approx \theta(\boldsymbol{d_{obs}}, \boldsymbol{m})$$

where $\sigma\,(\boldsymbol{d}, \boldsymbol{m})$ is the posterior distribution in the joint data-model parameter space $D \times M$, and $\theta(\boldsymbol{d}, \boldsymbol{m})$ is the distribution in $D \times M$ that describes the correlation between model parameters and data (allowing an uncertain forward relation). Since the true modelization error $\delta\boldsymbol{m}_{true}$ is unknown, we build an approximate modelization error distribution $\theta(\boldsymbol{d_{obs}}, \boldsymbol{m})$ by considering a simplified inverse problem that is similar to the original problem: First we solve the problem using a simplified pseudo-inverse $h$ which gives us the rough estimate $\tilde{\mathbf{m}} = h(\mathbf{d}_{\text{obs}})$. We now compute synthetic data from the estimate, using the correct forward $g$: $\mathbf{d}_{synt} = g(\tilde{\mathbf{m}})$, and invert this result again using $h$: $\tilde{\mathbf{m}}_2 = h(\mathbf{d}_{synt})$. If $\boldsymbol{m}_{true}$ and $\tilde{\mathbf{m}}$ are close in the model space, this back-and-forth process approximately reveals the modelization error, which can now be estimated as $\delta\boldsymbol{m}_{approx} = \tilde{\mathbf{m}}_2 - \tilde{\mathbf{m}}$ (Khoshkholgh et al. 2021). A simple modelization error distribution $\theta(\boldsymbol{d_{obs}}, \boldsymbol{m})$ can now be constructed as an isotropic Gaussian with mean $\tilde{\mathbf{m}}$ and with the components of $\delta\boldsymbol{m}_{approx}$ as standard deviations. This approximate modelization error distribution is used as a global proposal $q(\boldsymbol{m}'|\boldsymbol{m})$. In this way, our external knowledge and information about the target distribution can be injected into the problem through the proposal. Using the global proposal will not bias the problem or distort the result, but just accelerate the sampling process and leads to a more efficient algorithm (Khoshkholgh et al., 2021).

## Results: Seismic Inversion

A 2D seismic profile is chosen from the Tyra field in the Danish part of North see. There are two well logs located in the chosen area. By simple interpretation and from the well information, a velocity function was generated. This velocity model was considered as the center of the informed proposal distribution, modelization error distribution was calculated from it. The following steps were taken to create the informed proposal distribution:

a) Synthetics were computed from the reflectivity obtained from the center velocity model.
b) Synthetics were inverted with a simple linear inversion (deconvolution) and a second reflectivity model is obtained.
c) Modelization error for the layer boundaries were calculated by finding the difference between the first and the second reflectivity model. The modelization error was then turned into a modelization error envelope.
d) The center velocity model and the error envelope were used for the global proposal strategy in the IPMC method.

As previously explained, our strategy for generating prior models is inspired from Khoshkholgh et al. 2021, except for the fact that in this study fault lines are introduced into the model during perturbation. The probability of having fault structures in each point is obtained from the differential stress field and used in the sampling procedure. The prior probability is non-zero for piece-wise constant velocity models (Khoshkholgh et al.,2021). Overall, perturbing the prior model is a combination of warping or deforming the layer boundary shape, perturbing and changing the velocity value in each layer and introducing fault lines with regard to the fault and fracture probability.

Figure 9 shows a 2D seismic profile from the Tyra field that was used in this study. Figure 10 shows the center velocity model obtained by simple interpretation. The reflectivity modelization error estimation is shown in figure 11 and the modelization envelope is shown in figure 12. In figure 13 the prior fracture probability model, obtained from the stress field, is shown.

To start sampling the posterior probability density the center velocity model is chosen as the starting model. The maximum velocity perturbation is chosen as 1 percent of the maximum velocity and the maximum displacement in warping is chosen as 5 pixels. Warping happens in a square window centered at a randomly selected point. The size of the window is 200 by 200. The maximum and minimum fault angles are 12 degrees and 36 degrees respectively.

Figure 14 shows two different realizations from posterior. Using a global, informed proposal containing external information about the posterior sped up the sampling process, and made the sampling possible for the full model with more than $10^6$ parameters.

The posterior fault/fracture density is shown in figure 15 and suggests that the bottom of our 2D selected area, which is located under the reservoir, has a very low probability of having a fault or a fracture zone.

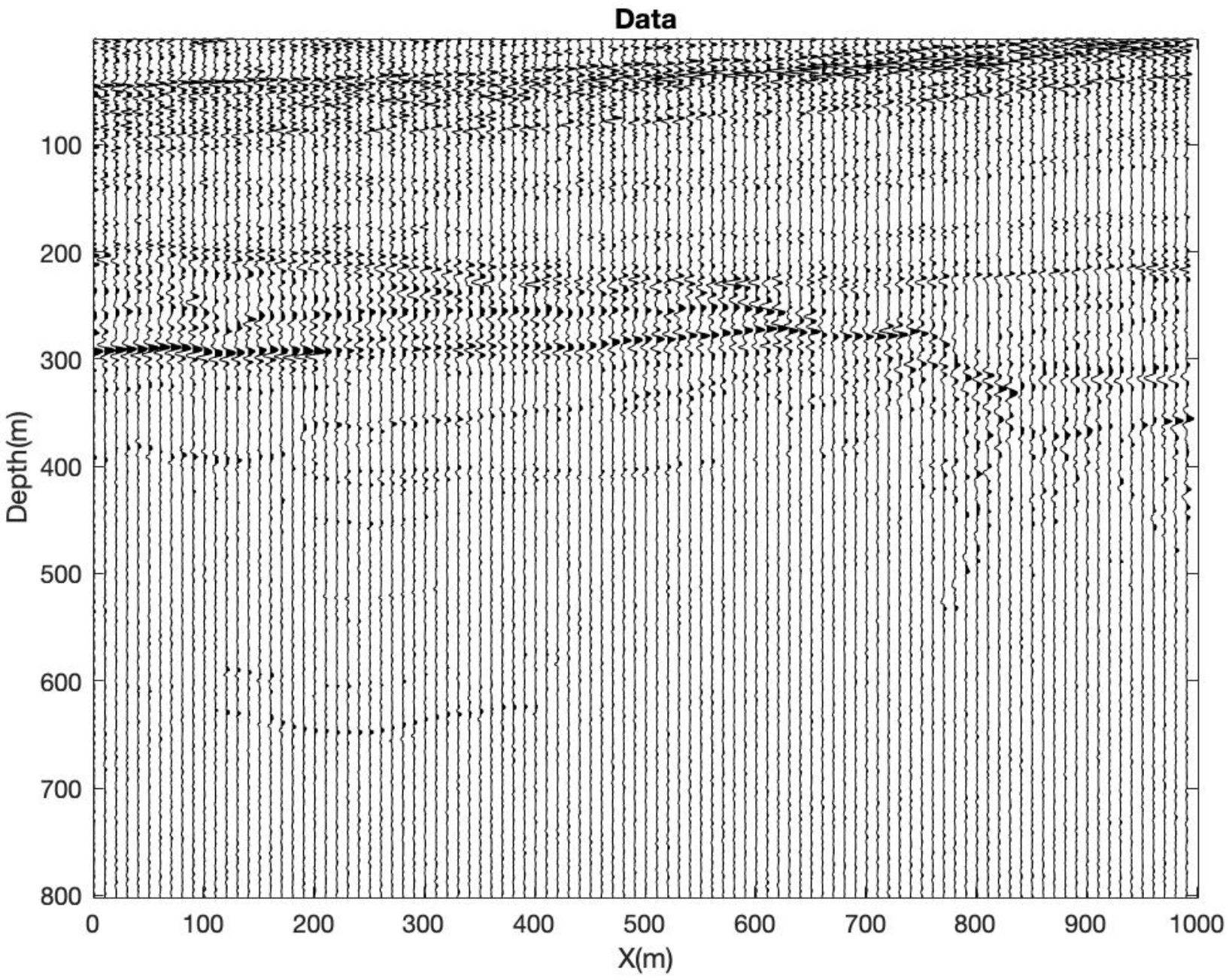

**Figure 9**. 2D observed seismic profile from the Tyra field.

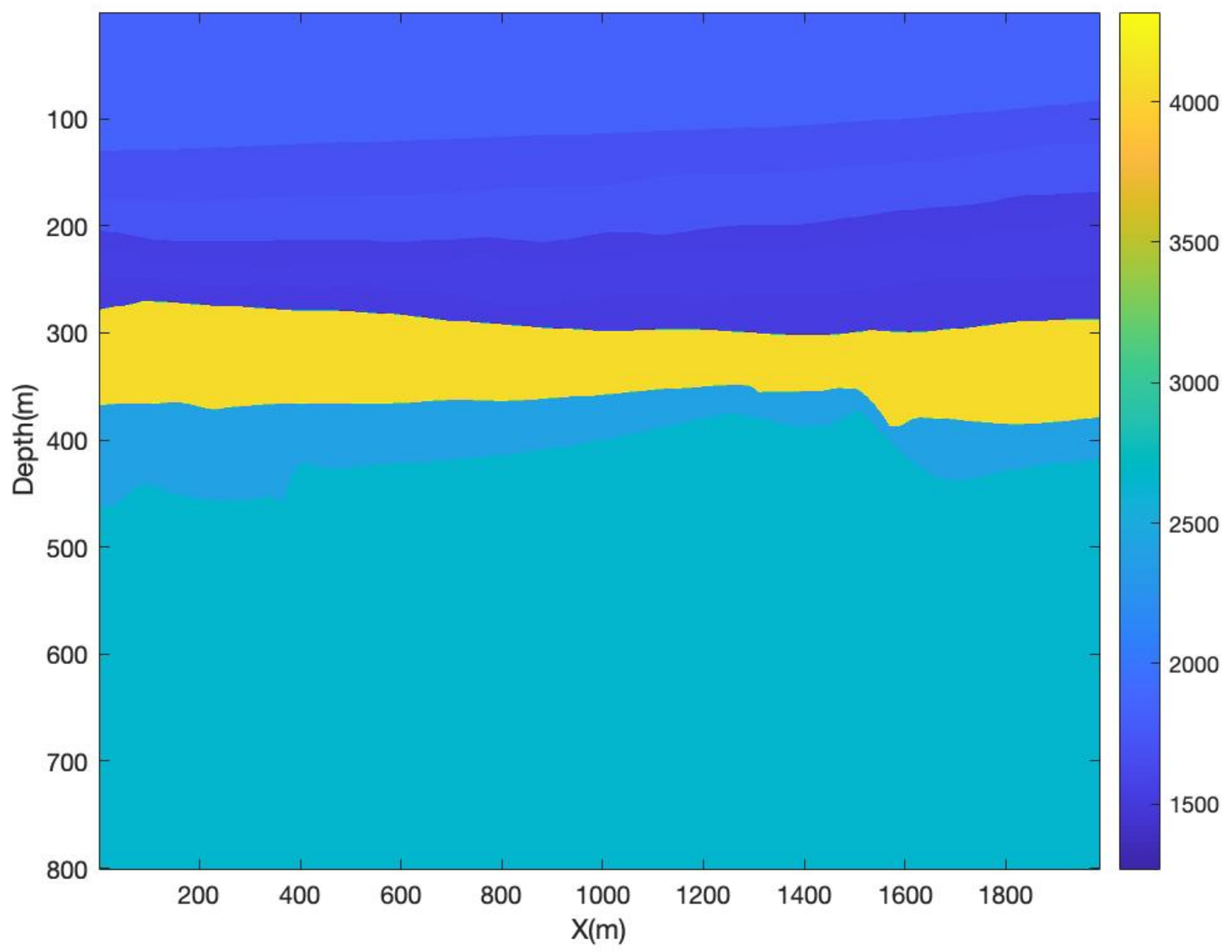

**Figure 10**. The center velocity model obtained by simple interpretation.

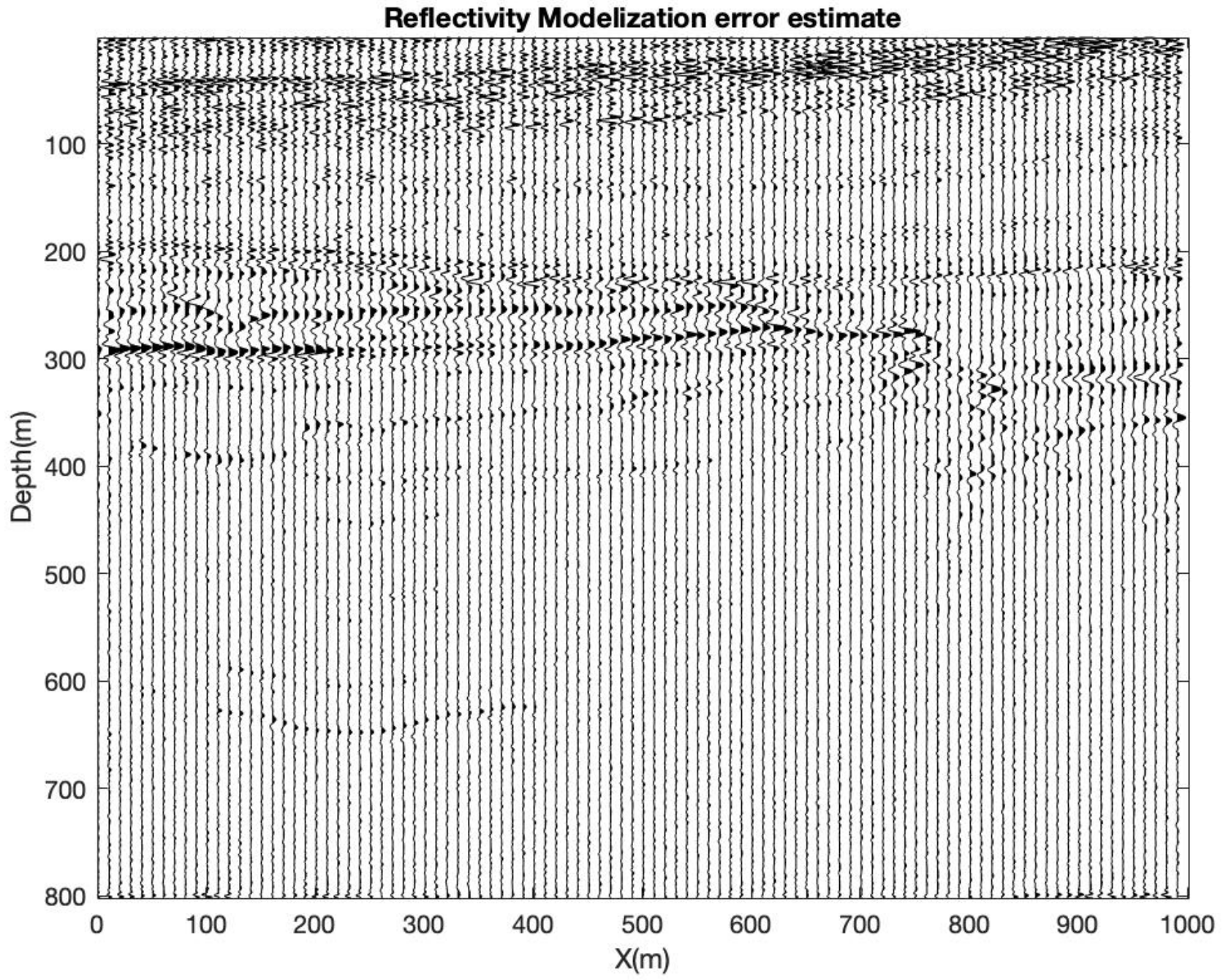

**Figure 11**. The reflectivity modelization error estimation.

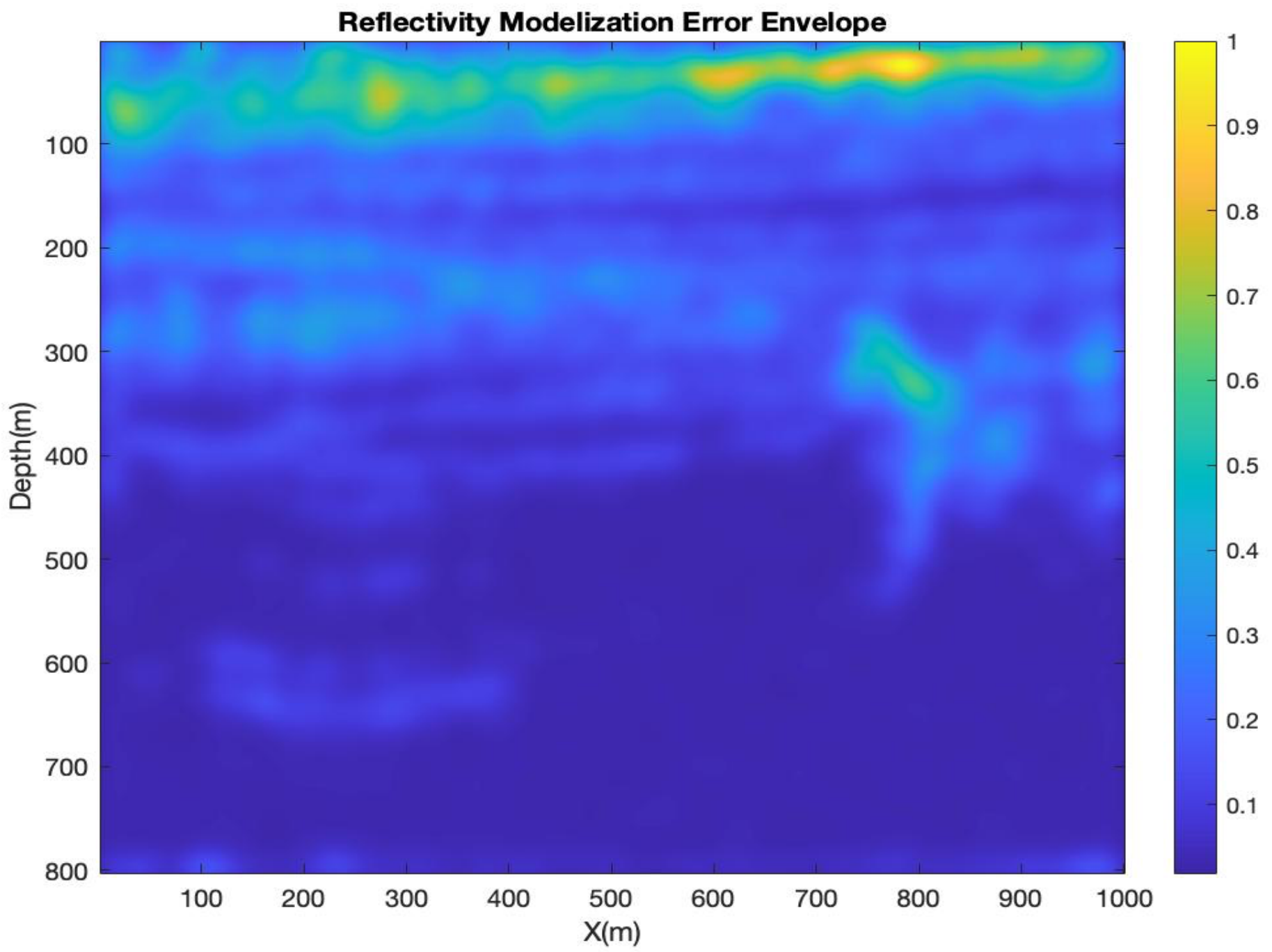

**Figure 12**. Reflectivity modelization error envelope.

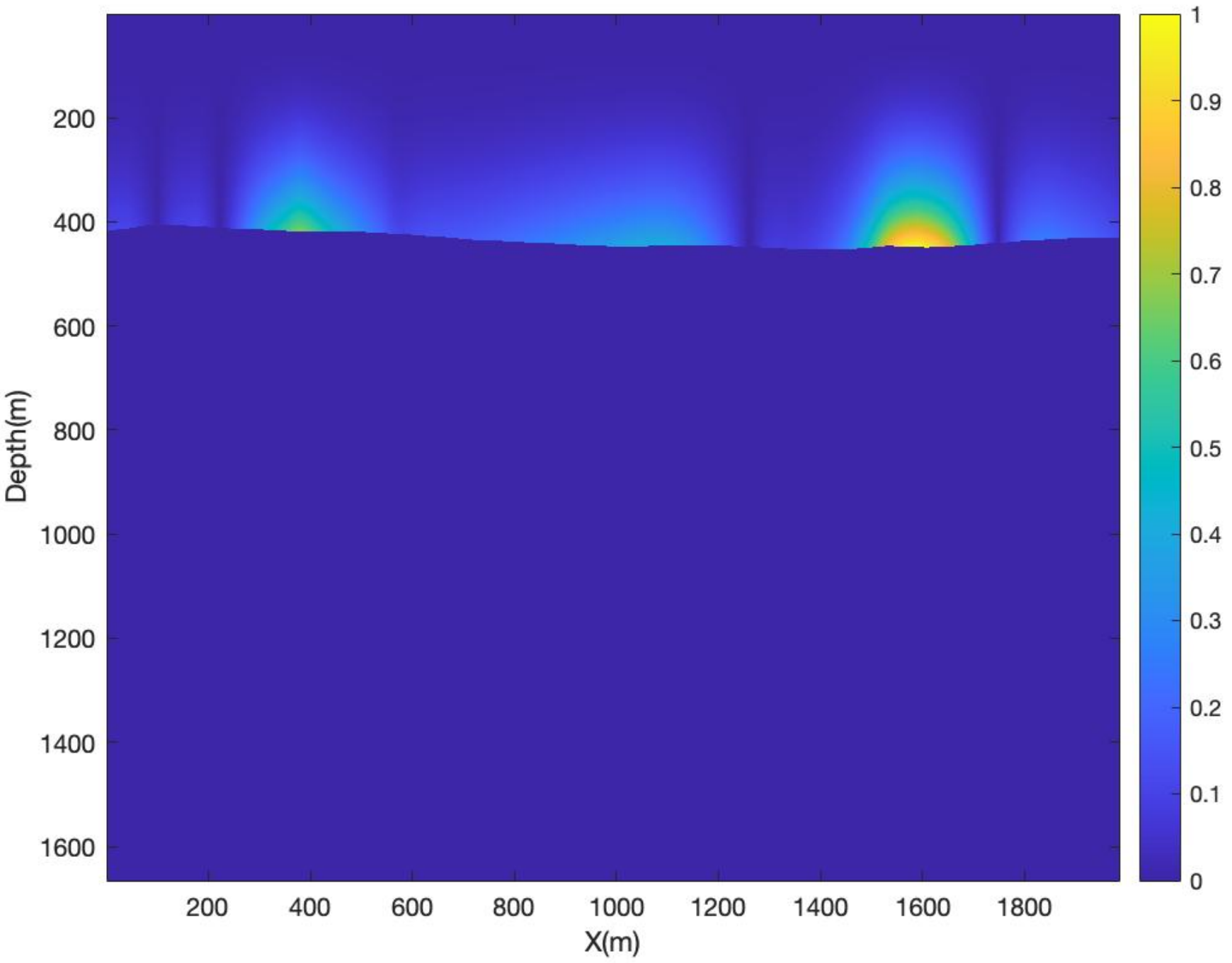

**Figure 13**. Prior fracture probability model obtained from the stress field.

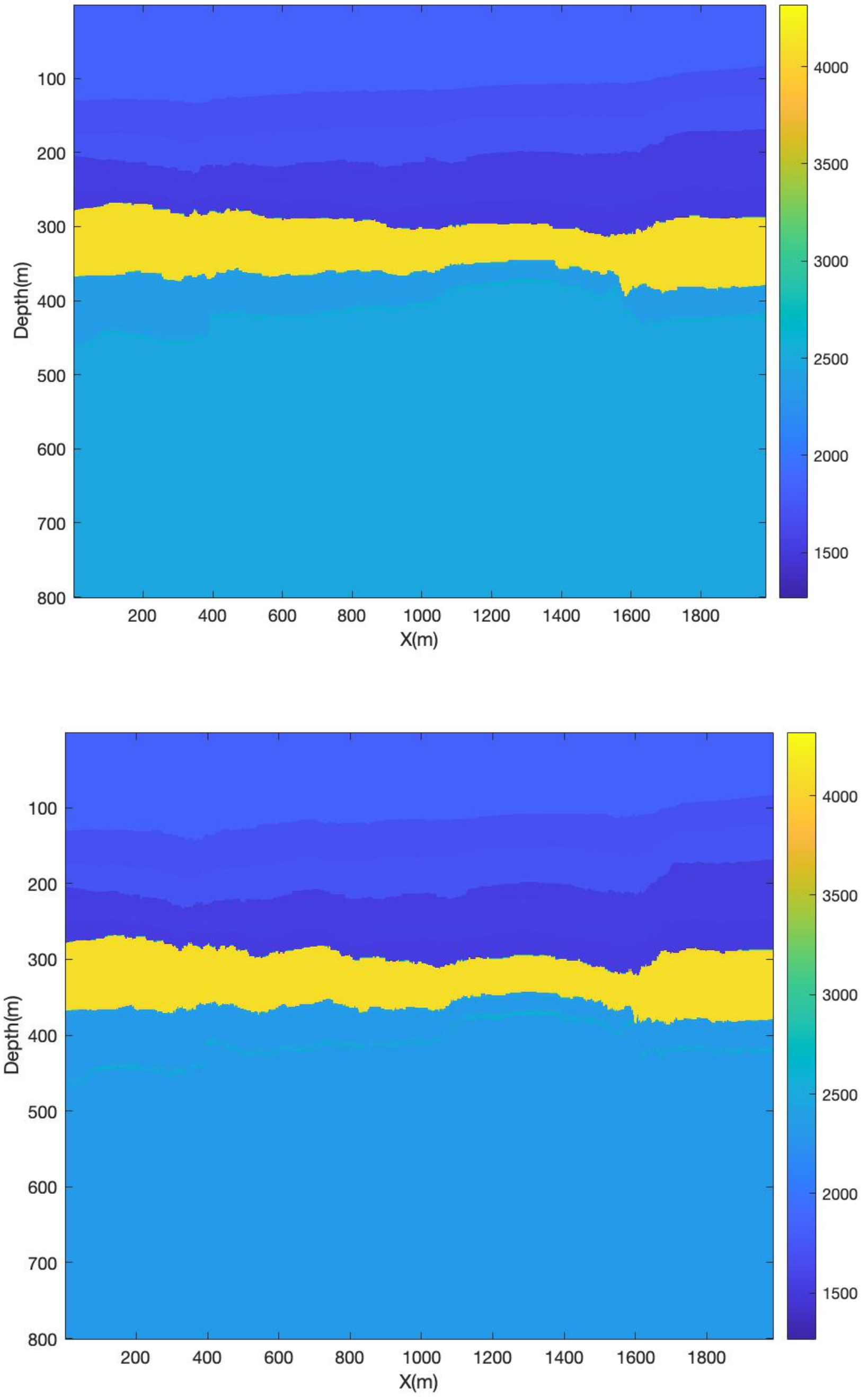

**Figure 14**. Two realizations from the posterior probability density.

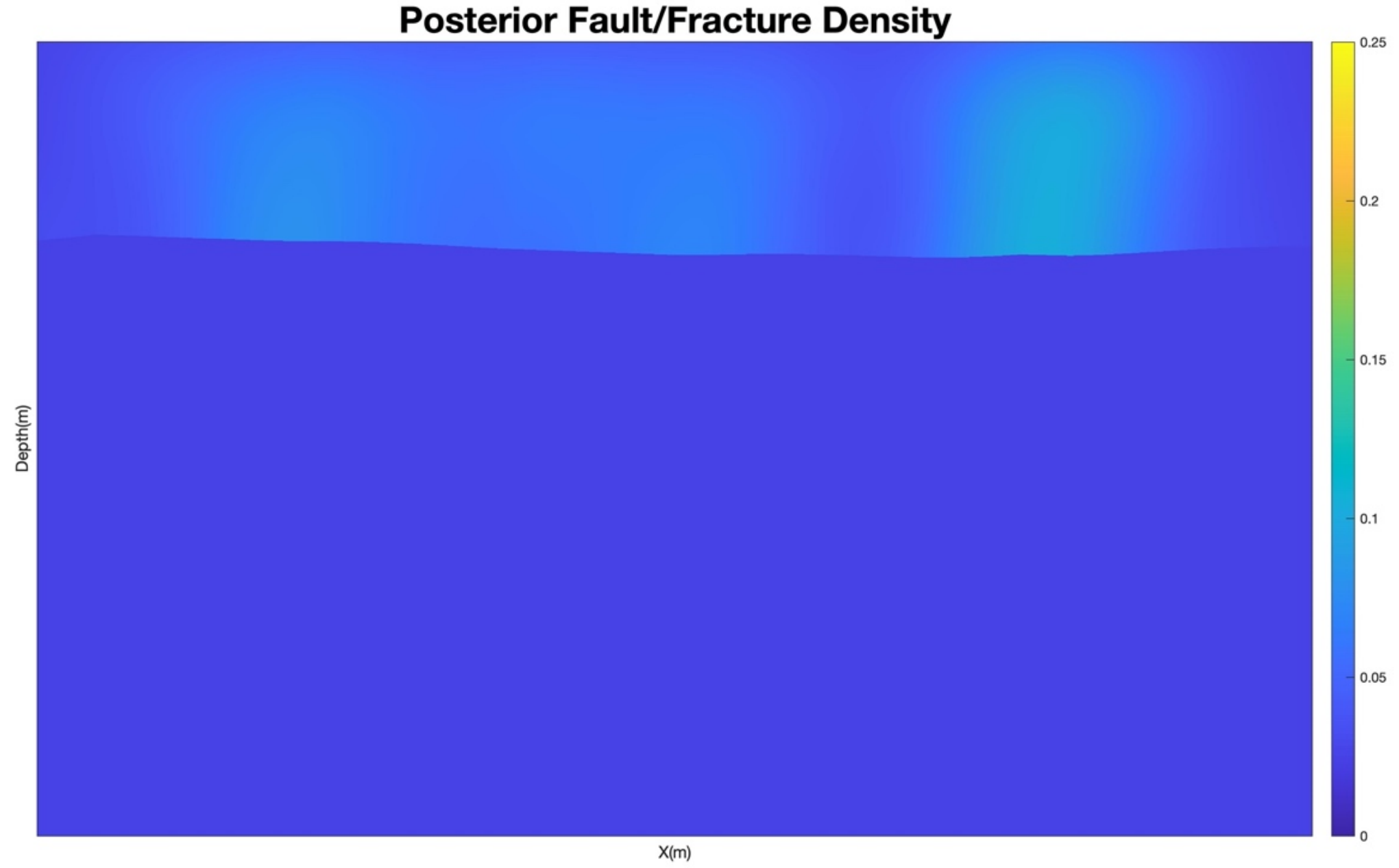

**Figure 15**. Posterior Fault and fracture density.

## Conclusions and Discussions

A forward conceptual model was used to simulate the evolution in time of the stress and strain fields in the Tyra field as result of the deposition of the approximately 2000m thick clay-rich Post-Chalk overburden in the last 61 Million years.

The main purpose was to study possible risks, in terms of probability of fracturing or fault re-activation, related to seal integrity.

Since the resolution of even the best quality seismic data does not allow to detect small fractures and faults, the forward conceptual model allowed us to identify areas of stress-strain concentration, which possibly can be at increased risk of fracturing and probable fault re-activation, especially in the cap rock (immediate overburden).

The simulation showed that the deposition of thick clay-rich Cenozoic Overburden in the last 61 MY on the top of a pre-existing structure could have produced significant shear stress in the strata below the Overburden, i.e. the Upper Jurassic, the Lower Cretaceous and the Chalk Group. The largest values of the shear stress were estimated along the very steep sections at Top Jurassic, which might corresponds to fault zones. The areas of high shear stress could be at higher risk for fracturing.

The largest effective strain (deformation) was estimated in the section above Top Chalk, which forms the seal of the Chalk reservoir.

The very intensive sedimentation in last 5.3 MY seems to lead to significant increase and re-distribution in both the shear stress of the deepest strata and the strain accumulated in the overburden.

The projection in time after end deposition showed that the strain-stress field continues to evolve even if no new sedimentation is occurring.

The stress-strain field, estimated with the present model is believed to correspond to natural state of the subsurface, i.e. the state before any kind of human intervention.

It should be kept in mind that the forward model of the evolution of the subsurface presented here is conceptual and was used mainly to illustrate the importance of the Cenozoic deposition for the present-day subsurface stress-strain field. The model can be improved by including further details.

A simplified probabilistic analysis of a seismic 2D profile across the Tyra reservoir, intersecting the E1-X and Deep Adda well sites was carried out. The focus was on the detection limit of the data, and the result was a map of faulting and fracturing probabilities across the area. The result is a combination of seismic information with information about stress fields, derived from the subsurface evolution model.

Based on this study, we currently cannot make very detailed conclusions about the actual conditions at specific locations in the Tyra field. Both the subsurface evolution model and the seismic analysis were oversimplified and important processes, among others, anisotropy in stress and rock properties, the pre-Cenozoic geological and tectonic history, the effect of temperature and capilliary phenomena were not included in the current study. A notable weakness in the seismic analysis is the lack of accurate data modelling (the wavelet estimation and the wave simulation), and we have also not included an obvious source of information, namely anisotropy of rock properties, which may be derived from full waveform data or AVO/AVA data. An investigation of anisotropy may contain valuable information about rock fracturing, and should be taken up in future studies.

The outcome of the study is, however, encouraging. We have proposed a new way of analysing caprock integrity of reservoirs for $CO_2$ storages. Our subsurface evolution model potentially allows us to predict current and future changes in a reservoir, and our probabilistic approach to seismic data analysis allows our numerical model of the stress field to be integrated with seismic information, producing fault/fracture probability maps that are ready to be included in a quantitative risk analysis.

## Acknowledgements

The authors would like to thank the DHRTC for providing the data the financial support for this study.

Ivanka Orozova-Bekkevold acknowledges the support of this work by Halliburton Software and Services, a Halliburton Company by allowing the authorized use of the ”Drillworks/Predict” software used to produce Figure 1 and 2.